\documentclass[journal, twocolumn, 10pt]{IEEEtran}
\usepackage{hyperref}
\usepackage{url}
\usepackage{graphicx}
\usepackage{epstopdf}
\usepackage{algorithm}
\usepackage{algorithmicx}
\usepackage{algpseudocode}
\usepackage{cite}
\usepackage{color}
\usepackage{amsmath,amssymb,amsfonts,amsthm}
\usepackage{bm}
\usepackage{setspace}
\usepackage{multirow}
\usepackage{mathtools}
\usepackage{wrapfig}

\begin{document}
\title{Phase Transitions in Spectral Community Detection}

\author{Pin-Yu~Chen and Alfred O. Hero III,~\emph{Fellow},~\emph{IEEE}
\\ Department of Electrical Engineering and Computer Science, University of Michigan, Ann Arbor, USA
\\Email : \{pinyu,hero\}@umich.edu
\thanks{This work has been partially supported by the Army Research Office (ARO), grant W911NF-09-1-0310, and by the Consortium for Verification Technology under Department of Energy National Nuclear Security Administration, award DE-NA0002534.}
}

\newcommand{\asconv}{\overset{\text{a.s.}}{\longrightarrow}}
\newcommand{\ra}{\rightarrow}
\newcommand{\LB}{\left\{}
\newcommand{\RB}{\right\}}
\newcommand{\Lb}{\left[}
\newcommand{\Rb}{\right]}
\newcommand{\lb}{\left(}
\newcommand{\rb}{\right)}
\newcommand{\cone}{c_1}
\newcommand{\ctwo}{c_2}
\newcommand{\pone}{p_1}
\newcommand{\ptwo}{p_2}
\newcommand{\none}{{n_1}}
\newcommand{\ntwo}{{n_2}}
\newcommand{\hnone}{\widehat{n}_1}
\newcommand{\hntwo}{\widehat{n}_2}
\newcommand{\Aone}{\mathbf{A}_1}
\newcommand{\Atwo}{\mathbf{A}_2}
\newcommand{\bDelta}{\mathbf{\Delta}}
\newcommand{\bA}{\mathbf{A}}
\newcommand{\bD}{\mathbf{D}}
\newcommand{\bL}{\mathbf{L}}
\newcommand{\bC}{\mathbf{C}}
\newcommand{\bCbar}{\mathbf{\bar{C}}}
\newcommand{\bAbar}{\mathbf{\bar{A}}}
\newcommand{\bDbar}{\mathbf{\bar{D}}}
\newcommand{\bone}{\mathbf{1}}
\newcommand{\onenone}{\mathbf{1}_{\none}}
\newcommand{\onentwo}{\mathbf{1}_{\ntwo}}
\newcommand{\Done}{\mathbf{D}_1}
\newcommand{\Dtwo}{\mathbf{D}_2}
\newcommand{\Lone}{\mathbf{L}_1}
\newcommand{\Ltwo}{\mathbf{L}_2}
\newcommand{\yone}{\mathbf{y}_1}
\newcommand{\ytwo}{\mathbf{y}_2}
\newcommand{\xone}{\mathbf{x}_1}
\newcommand{\xtwo}{\mathbf{x}_2}
\newcommand{\bx}{\mathbf{x}}
\newcommand{\by}{\mathbf{y}}
\newcommand{\bz}{\mathbf{z}}
\newcommand{\pUB}{p_{\text{UB}}}
\newcommand{\pLB}{p_{\text{LB}}}
\newcommand{\hpUB}{\widehat{p}_{\text{UB}}}
\newcommand{\hpLB}{\widehat{p}_{\text{LB}}}
\newcommand{\hLone}{\widehat{\mathbf{L}}_1}
\newcommand{\hLtwo}{\widehat{\mathbf{L}}_2}
\pagestyle {empty}
\thispagestyle{empty}

\maketitle
\thispagestyle{empty}
\begin{abstract}
Consider a network consisting of two subnetworks (communities) connected by some external edges. Given the network topology,
the community detection problem can be cast as a graph partitioning problem that aims to identify the external edges as the graph cut that separates these two subnetworks. In this paper, we consider a general model where two arbitrarily connected subnetworks are connected by random external edges. Using random matrix theory and concentration inequalities, we show that when one performs community detection via spectral clustering there exists an abrupt phase transition as a function of the random external edge connection probability. Specifically, the community detection performance transitions from almost perfect detectability to low detectability near some critical value of the random external edge connection probability. We derive upper and lower bounds on the critical value and show that the bounds are equal to each other when two subnetwork sizes are identical. Using simulated and experimental data we show how these bounds can be empirically estimated to validate the detection reliability of any discovered communities.
\end{abstract}


\section{Introduction}
\label{sec_Intro}
Recently, graph signal processing has been an active research field in data processing and inference \cite{Sandryhaila13,Bertrand13,Shuman13,Dong14,MillerICASSP10,SihengChen14}. Community detection \cite{Fortunato10} is a typical example of graph signal processing where the signal is a graph representing connectivity structure and the goal is to identify communities from the graph. 
Applications of community detection include data clustering \cite{Luxburg07}, 
social and biological network analysis \cite{Girvan02,Newman06PNAS}, and network vulnerability assessment \cite{CPY13GlobalSIP,CPY14ComMag}, among others.
This paper provides fundamental limits affecting community detectability for spectral clustering methods. These limits are in the form   of a phase transition threshold in the algebraic connectivity of the network as a function of the random inter-community edge connection probability.		

Consider a network consisting of two node-disjoint subnetworks (communities) connected by some external edges. Let $n$ denote the total number of nodes in a network. For an undirected and unweighted graph, the network topology can be characterized by its adjacency matrix $\bA$, where $\bA$ is a binary symmetric $n \times n$ matrix, with $\bA_{ij}=1$ if an edge exists between node $i$ and node $j$, and $\bA_{ij}=0$ otherwise.
Given the adjacency matrix of the entire network,
the community detection problem can be cast as a graph partitioning problem that identifies the external edges as the graph cut that separates these two subnetworks. It also can be viewed as a clustering problem when the input data is a binary adjacency matrix (e.g., friendship graph) instead of a similarity graph.
Note that in practice finding such a graph cut is a nontrivial task since the computational complexity of graph cut algorithms is high for large dense networks.

Several authors \cite{White05,Gennip12,Tsironis13,Huang14} have proposed to use spectral clustering \cite{Luxburg07,Shi00} for community detection, but the  detectability of spectral clustering is poorly understood.
 Spectral clustering specifies a graph cut by inspecting the eigenstructure of a graph. Let $\mathbf{1}_n~(\mathbf{0}_n)$ denote the all-one (all-zero) vector of length $n$ and let $\bD=\text{diag}(\bA \mathbf{1}_n)$ be a diagonal matrix with degree information on its main diagonal. Define the graph Laplacian matrix as $\bL=\bD-\bA$. Let $\lambda_i(\bL)$ be the $i$th smallest eigenvalue of $\bL$. It is well-known that $\lambda_1(\bL)=0$ since $\bL \mathbf{1}_n=\mathbf{0}_n $ and $\bL$ is a positive semidefinite (PSD) matrix \cite{Merris94,Chung97SpectralGraph}. The second smallest eigenvalue, $\lambda_2(\bL)$, is known as the algebraic connectivity. $\lambda_2(\bL) >0$ if and only if the network is a connected graph. The eigenvector associated with $\lambda_2(\bL)$, denoted by $\by$, is also called the Fiedler vector \cite{Fiedler73}. A mathematical representation of the algebraic connectivity is
\begin{align}
\label{eqn_alge}
\lambda_2(\bL)= \min_{\|\bx\|_2=1,\mathbf{1}_n^T \bx=0} \bx^T \bL \bx.
\end{align}

The spectral clustering method for community detection \cite{White05,Gennip12,Tsironis13,Huang14} is summarized as follows:\\
(1)~~Compute the graph Laplacian matrix $\mathbf{L}=\mathbf{D}-\mathbf{A}$.\\
(2)~~Compute the Fiedler vector $\by$.\\
(3)~~Perform K-means clustering on $\by$ to cluster the nodes into two groups.\\
Note that K-means clustering determines two centroids based on the Fiedler vector $\by$ and then label each node to the closest cluster
according to the Euclidean distances of its corresponding entry in $\by$ to the centroids. The graph cut is the set of edges between the two identified communities.
For community detection on more than two subnetworks, we can use successive spectral clustering on the discovered subnetworks.

In this paper, we establish the existence of an abrupt phase transition for community detection based on spectral clustering. At some critical value of random external connection probability, the network transitions from one admitting almost perfect detectability to one in which the subnetworks cannot be identified accurately. We provide upper and lower bounds on this critical value. The bounds become equal to each other, yielding an exact expression for the critical value, when these two subnetwork sizes are identical. This framework can be generalized to community detection on more than two subnetworks by aggregating multiple subnetworks into two larger subnetworks. We show how these bounds can be empirically estimated from real network data in order to validate that the detector is operating in a regime where community detection is reliable, i.e., below the phase transition threshold.


\section{Related Works}
Community detection arises in technological, social, and biological networks. For social science, the goal is to find tightly connected subgraphs in a social network \cite{Fortunato10}. In \cite{Newman06PNAS}, Newman proposes a measure called modularity that evaluates the number of excessive edges of a graph compared with the corresponding degree-equivalent random graph. More specifically, define the modularity matrix as $\mathbf{B}=\bA-\frac{\mathbf{d}\mathbf{d}^T}{2m}$, where $\mathbf{d}=\bA \mathbf{1}_n$ is the degree vector and $m$ is the number of edges in the graph. The last term $\frac{\mathbf{d}\mathbf{d}^T}{2m}$ is the expected adjacency matrix of the degree-equivalent random graph. Similar to spectral clustering, the community indication vector is obtained by performing K-means clustering on the largest eigenvector of $\mathbf{B}$. We will compare the community detection results of spectral clustering and the modularity method in Sec. \ref{Sec_performance}.

The stochastic block model \cite{Holland83} is widely used for community detection as it parameterizes community detection problems with a small number of parameters \cite{Karrer11}, where the parameters are common edge connection probabilities within and between each subnetwork. Furthermore, it has been shown that the stochastic block model can provide a good fit to real-world community data \cite{Karrer11,Olhede14}.
Under the stochastic block model, many authors have observed apparent phase transition phenomenon on community detectability for different community detection algorithms when one gradually increases the number of external edges between communities \cite{Bickel09,Ronhovde09,Krzakala13,Radicchi14}.
The detectability of the modularity method is studied in \cite{Nadakuditi12Detecability} when the two subnetworks are of equal size and each node pair in each subnetwork is randomly connected by the same edge connection probability. The planted clique detection problem in \cite{Nadakuditi12plant} is a further restriction of the stochastic block model. In \cite{Decelle11}, the authors study phase transitions on community detectability for sparse random networks generated by the stochastic block model. A universal phase transition threshold on community detectability for the modularity method under the stochastic block model is established in \cite{CPY14modularity}, where
the asymptotic critical value depends only on the parameters of the stochastic block model and does not depend on the ratio of community sizes in the large network limit.

Our model is more general than the stochastic block model since it does not assume any edge connection models within the communities. The details are discussed in Sec. \ref{sec_system}. A similar model is studied in \cite{Radicchi13} for interconnected networks. However, in \cite{Radicchi13} the subnetworks are of equal size and the external edges are known (i.e., non-random). The main contribution of \cite{Radicchi13} was a study of the eigenstructure of the overall graph Laplacian matrix with different interconnected edge strengths as contrasted to community detection. The simulation results in \cite{Radicchi13_hetero} show that phase transition on community detectability exists under this general model, yet the critical phase transition threshold is still poorly understood.
Phase transition results on p-resistance distances of random geometric graphs are obtained in \cite{Alamgirv11}. The authors of \cite{Alamgirv11} show that there exist two critical thresholds for the p-resistance. The first (lower) threshold depends on the global graph topology while the second (higher) threshold only depends on local graph connectivity.

\section{Network Model and Phase Transitions in Spectral Community Detection}
\label{sec_system}
Consider two arbitrarily connected subnetworks with internal adjacency matrices $\Aone$ and $\Atwo$ and network sizes $\none$ and $\ntwo$, respectively. The external connections between these two subnetworks are characterized by a binary $\none \times \ntwo$ adjacency matrix $\bC$.
We assume Erdos-Renyi random model for external edges, where each entry in $\bC$ is a Bernoulli($p$) random variable.
Let $n=\none+\ntwo$. The overall $n \times n$ adjacency matrix can be represented as

\begin{align}
\label{eqn_asym_block_model}
\mathbf{A} = \begin{bmatrix}
       \Aone & \bC           \\
       \bC^T           & \Atwo
     \end{bmatrix}.
\end{align}
The network model (\ref{eqn_asym_block_model}) is very general as it does not impose restrictive conditions on the forms of $\Aone$ and $\Atwo$.
The two subnetworks can have arbitrary network structures as long as each subnetwork is connected.
Therefore, the proposed model (\ref{eqn_asym_block_model}) fits any stochastic model for community structure that has constant inter-community connectivity parameters. For example, (\ref{eqn_asym_block_model}) is equivalent to a stochastic block model given stochastic realizations of the subnetwork adjacency matrices $\Aone$ and $\Atwo$. In the
stochastic block model the two subnetworks are assumed to be generated by Erdos-Renyi graphs, i.e., the internal connections are governed by constant subnetwork-wide connection probability between each node pair.
Specifically, the stochastic block model is specified by a $2\times 2$ connection probability matrix
\begin{align}
\label{eqn_group_connection_matrix}
\mathbf{P} = \bordermatrix{~ & \text{subnetwork 1} & \text{subnetwork 2} \cr
                  \text{subnetwork 1} & \pone & p \cr
                  \text{subnetwork 2} & p & \ptwo \cr},
\end{align}
where $p_i$ is the internal edge connection probability for subnetwork $i$.
Thus the adjacency matrix $\bA_i$ in (\ref{eqn_asym_block_model}) can be interpreted as a connected realization of a Erdos-Renyi graph with edge connection probability $p_i$.
The planted clique detection problem in \cite{Nadakuditi12plant} is a special case of (\ref{eqn_group_connection_matrix}) when $p_1=1$ and $p_2=p$. The analysis below holds for random graph distributions that are more general than the stochastic block model. We only need to assume that the connections between the two arbitrarily connected subnetworks are random with probability $p$.
Thus, the phase transition results obtained in Sec. \ref{sec_bounds} hold for
the stochastic block model (\ref{eqn_group_connection_matrix}), and indeed for any stochastic model of intra-community connectivity, for any $\pone, \ptwo >0$.

Let $\onenone$ be the all-one vector of length $\none$ and $\onentwo$ be the all-one vector of length $\ntwo$, and let $\Done=\text{diag}\left(\bC\onentwo\right)$ and $\Dtwo=\text{diag}\left(\bC^T\onenone\right)$.
The corresponding overall graph Laplacian matrix can be represented as
\begin{align}
\label{eqn_Laplacian_block}
\mathbf{L} = \begin{bmatrix}
       \Lone+\Done & -\bC           \\
       -\bC^T           & \Ltwo+\Dtwo
     \end{bmatrix},
\end{align}
where $\Lone$ and $\Ltwo$ are the graph Laplacian matrices of subnetworks 1 and 2, respectively. Let $\mathbf{x}=[\xone~\xtwo]^T$, where $\xone \in \mathbb{R}^\none$ and $\xtwo \in \mathbb{R}^\ntwo$. By (\ref{eqn_alge}) we have
$\lambda_2(\mathbf{L})=\min_{\bx} \bx^T \mathbf{L} \bx$ subject to the constraints $\xone^T\xone+\xtwo^T\xtwo=1$ and $\xone^T\onenone+\xtwo^T\onentwo=0$. Using Lagrange multipliers $\mu$, $\nu$ and (\ref{eqn_Laplacian_block}), the Fiedler vector $\by=[\yone~\ytwo]^T$ of $\mathbf{L}$, with $\yone \in \mathbb{R}^\none$ and $\yone \in \mathbb{R}^\ntwo$, satisfies
$\by=\arg \min_{\bx} \Gamma(\bx)$, where
\begin{align}
\label{eqn_Lagrange}
\Gamma(\bx)&=\xone^T (\Lone+\Done) \xone + \xtwo^T (\Ltwo+\Dtwo) \xtwo -2\xone^T \bC \xtwo \nonumber \\
 &~~~-\mu(\xone^T\xone+\xtwo^T\xtwo-1)-\nu (\xone^T\onenone+\xtwo^T\onentwo).
\end{align}
Differentiating (\ref{eqn_Lagrange}) with respect to $\xone$ and $\xtwo$ respectively, and substituting $\by$ into the equations, we obtain
\begin{align}
\label{eqn_Lagrange1}
&2(\Lone+\Done) \yone -2\bC \ytwo - 2\mu \yone -\nu \onenone=\mathbf{0}_{\none}, \\
\label{eqn_Lagrange2}
&2(\Ltwo+\Dtwo) \ytwo -2\bC^T \yone - 2\mu \ytwo -\nu \onentwo=\mathbf{0}_{\ntwo}.
\end{align}
Left multiplying (\ref{eqn_Lagrange1}) by $\onenone^T$ and left multiplying (\ref{eqn_Lagrange2}) by $\onentwo^T$ , we have
\begin{align}
\label{eqn_Lagrange3}
&2\onenone^T\Done \yone -2\onenone^T\bC \ytwo - 2\mu \onenone^T\yone -\nu \none=0, \\
\label{eqn_Lagrange4}
&2\onentwo^T\Dtwo \ytwo -2\onentwo^T\bC^T \yone - 2\mu \onentwo^T\ytwo -\nu \ntwo=0.
\end{align}
Since $\onenone^T\Done=\onentwo^T\bC^T$ and $\onenone^T\bC=\onentwo^T\Dtwo$, adding (\ref{eqn_Lagrange3}) and (\ref{eqn_Lagrange4}) together we obtain
$\nu=-\frac{2\mu}{n} (\yone^T\onenone+\ytwo^T\onentwo)$,
which is equivalent to $0$ since $\bone_n^T \by=0$ as $\by$ is the Fiedler vector.
Applying $\nu=0$ and left multiplying  (\ref{eqn_Lagrange1}) by $\yone^T$ and left multiplying (\ref{eqn_Lagrange2}) by $\ytwo^T$, we have
\begin{align}
\label{eqn_Lagrange5}
&\yone^T (\Lone+\Done) \yone -\yone^T \bC \ytwo-\mu \yone^T \yone =0,\\
\label{eqn_Lagrange6}
&\ytwo^T (\Ltwo+\Dtwo) \ytwo -\ytwo^T \bC^T \yone-\mu \ytwo^T \ytwo =0.
\end{align}
Adding them together and by (\ref{eqn_alge}) and (\ref{eqn_Laplacian_block}) we obtain $\mu=\lambda_2(\mathbf{L})$.

Let $\bCbar=p \onenone \onentwo^T$, a matrix whose elements are the means of entries in $\bC$. Let $\sigma_i(\mathbf{M})$ denote the $i$th largest singular value of $\mathbf{M}$
\footnote{Note that for convenience, we use $\lambda_i({\mathbf{M}}_1)$ to denote the $i$th smallest eigenvalue of a square matrix $\mathbf{M}_1$ and use $\sigma_i({\mathbf{M}}_2)$ to denote the $i$th largest singular value of a rectangular matrix $\mathbf{M}_2$.}
 and write
$\bC=\bCbar+\bDelta$,
where $\bDelta=\bC-\bCbar$. By Latala's theorem \cite{Latala05},
\begin{align}
\label{eqn_Latala}
\mathbb{E} \Lb \sigma_1\lb \frac{\bDelta}{\sqrt{\none \ntwo}} \rb \Rb \rightarrow 0.
\end{align}
This is proved in Appendix \ref{appen_Latala}.
Furthermore, by Talagrand's concentration inequality \cite{Talagrand95},
\begin{align}
\label{eqn_Talagrand}
\sigma_1\lb \frac{\bC}{\sqrt{\none \ntwo}} \rb \asconv p
\text{~~and~~}
\sigma_i\lb \frac{\bC}{\sqrt{\none \ntwo}} \rb \asconv 0~~\forall i \geq 2
\end{align}
when $\none \rightarrow \infty$ and $\ntwo \rightarrow \infty$, and $\asconv$ denotes almost sure convergence. This is proved in Appendix \ref{appen_Talagrand}. Note that the convergence rate is maximal when $\none=\ntwo$ because $\none+\ntwo \geq 2\sqrt{\none \ntwo}$ and the equality holds if $\none=\ntwo$.
The interpretation is that the convergence rate is governed by the subnetwork with the smallest size.
Throughout this paper we further assume $\frac{\none}{\ntwo} \ra c >0$ as $\none,\ntwo \ra \infty$. This means the subnetwork sizes grow with comparable rates.

As proved in \cite{BenaychGeorges12}, the singular vectors of $\bC$ and $\bCbar$ are close to each other in the sense that the square of inner product of their left/right singular vectors converges to $1$ almost surely when $\sqrt{\none \ntwo} p \rightarrow \infty$.
Consequently, we have
\begin{align}
\label{eqn_D1}
&\frac{1}{\ntwo}\Done \onenone=\frac{1}{\ntwo}\bC \onentwo \asconv p \onenone; \\
\label{eqn_D2}
&\frac{1}{\none}\Dtwo \onentwo=\frac{1}{\none}\bC^T \onenone \asconv p \onentwo.
\end{align}

Applying (\ref{eqn_Talagrand}), (\ref{eqn_D1}) and (\ref{eqn_D2}) to (\ref{eqn_Lagrange3}) and (\ref{eqn_Lagrange4}) and recalling that $\nu=0$ and $\frac{\none}{\ntwo}=c>0$,
 we have
\begin{align}
\label{eqn_Lagrange7}
&\frac{1}{\sqrt{c}}p \onenone^T \yone-\sqrt{c} p \onentwo^T \ytwo - \frac{1}{\sqrt{\none\ntwo}}\mu \onenone^T \yone \asconv 0;  \\
\label{eqn_Lagrange8}
&\sqrt{c} p \onentwo^T \ytwo-\frac{1}{\sqrt{c}} p \onenone^T \yone - \frac{1}{\sqrt{\none\ntwo}}\mu \onentwo^T \ytwo \asconv 0.
\end{align}
By the fact that $\onenone^T \yone+\onentwo^T \ytwo=0$, we have
\begin{align}
\label{eqn_Lagrange9}
&\lb \sqrt{c} + \frac{1}{\sqrt{c}} \rb \left(p-\frac{\mu}{n} \right) \onenone^T \yone \asconv 0; \\
&\lb \sqrt{c} + \frac{1}{\sqrt{c}} \rb \left(p-\frac{\mu}{n} \right) \onentwo^T \ytwo \asconv 0.
\end{align}
Consequently, as $\mu=\lambda_2(\bL)$, at least one of the two cases have to be satisfied:
\begin{align}
\label{eqn_Lagrange11}
& \text{Case 1:}~\frac{\lambda_2(\mathbf{L})}{n}\overset{\text{a.s.}}{\longrightarrow}p. \\
\label{eqn_Lagrange12}
& \text{Case 2:}~\onenone^T \yone \asconv 0~~\text{and}~~\onentwo^T \ytwo \asconv 0.
\end{align}

The algebraic connectivity and the Fiedler vector $\by$ undergo a phase transition between Case 1 and Case 2 as a function of $p\in [0,1]$. That is, a transition from Case 1 to Case 2 occurs when $p$ exceeds a certain threshold $p^*$.
In Case 1, the asymptotic algebraic connectivity grows linearly with $p$ while the asymptotic Fiedler vector remains the same (unique up to its sign). Furthermore, from (\ref{eqn_Lagrange5}), (\ref{eqn_Lagrange6}), (\ref{eqn_Talagrand}), (\ref{eqn_Lagrange11}), $\mu=\lambda_2(\bL)$ and $\onenone^T \yone+\onentwo^T \ytwo=0$, the Fielder vector $\by$ in Case 1 has the following property.
\begin{align}
\label{eqn_Lagrange13}
&\frac{1}{\sqrt{\none \ntwo}}\yone^T \Lone \yone + \frac{p}{\sqrt{\none \ntwo}}(\onenone^T \yone)^2-\sqrt{c} p\yone^T\yone \asconv 0,\\
\label{eqn_Lagrange14}
&\frac{1}{\sqrt{\none \ntwo}}\ytwo^T \Ltwo \ytwo + \frac{p}{\sqrt{\none \ntwo}}(\onenone^T \yone)^2-\frac{1}{\sqrt{c}} p \ytwo^T\ytwo \asconv 0.
\end{align}
Adding (\ref{eqn_Lagrange13}) and (\ref{eqn_Lagrange14}), we have
\begin{align}
\label{eqn_Lagrange15}
&\frac{1}{\sqrt{\none \ntwo}}\left(\yone^T \Lone \yone + \ytwo^T \Ltwo \ytwo \right) + \nonumber \\
&\Lb \frac{2(\onenone^T \yone)^2}{\sqrt{\none \ntwo}}-\left(\sqrt{c} \yone^T\yone+\frac{1}{\sqrt{c}}\ytwo^T\ytwo\right) \Rb p
\overset{\text{a.s.}}{\longrightarrow}0.
\end{align}
As the parenthesized and bracketed terms in (\ref{eqn_Lagrange15}) converge to finite constants for all $p$ in Case 1,
\begin{align}
\label{eqn_Lagrange16}
&\frac{1}{\sqrt{\none \ntwo}}\left(\yone^T \Lone \yone + \ytwo^T \Ltwo \ytwo \right) \asconv 0; \\
&\frac{2(\onenone^T \yone)^2}{\sqrt{\none \ntwo}}-\left(\sqrt{c} \yone^T\yone+\frac{1}{\sqrt{c}}\ytwo^T\ytwo\right) \asconv 0.
\end{align}
By the PSD property of the graph Laplacian matrix, $\yone^T \Lone \yone>0$ and $\ytwo^T \Ltwo \ytwo>0$ if and only if $\yone$ and $\ytwo$ are not constant vectors. Therefore (\ref{eqn_Lagrange16}) implies $\yone$ and $\ytwo$ converge to constant vectors. By the constraints $\yone^T\yone+\ytwo^T\ytwo=1$ and $\onenone^T\yone+\onentwo^T\ytwo=0$, we have
\begin{align}
\label{eqn_Lagrange17}
\sqrt{\frac{n \none}{\ntwo}} \yone \asconv \pm \onenone
~~\text{and}~~
\sqrt{\frac{n \ntwo}{\none}} \ytwo \asconv \mp \onentwo.
\end{align}
Consequently, in Case 1 $\yone$ and $\ytwo$ tend to be constant vectors with opposite signs.

More importantly, these results suggest a phase transition effect in spectral clustering.
By (\ref{eqn_Lagrange17}) and the constraint that $\onenone^T \yone+\onentwo^T \ytwo=0$, we know that in Case 1, $\onenone^T \yone=-\onentwo^T \ytwo$, and the two centroids found by K-means clustering of step (3) in Sec. \ref{sec_Intro} will have opposite signs since $\left|\onenone^T \yone\right|=\left|\onentwo^T \ytwo\right| \neq 0$ almost surely. Therefore in Case 1
spectral clustering can almost correctly identify these two subnetworks since $\yone$ and $\ytwo$ are constant vectors with opposite signs. On the other hand, in Case 2, $\onenone^T \yone \ra 0$ and $\onentwo^T \ytwo \ra 0$ almost surely. The entries of $\yone$ and $\ytwo$
tend to have opposite signs within each subnetwork. Therefore, in Case 2 spectral clustering leads to very poor community detection.

\section{Upper and Lower Bounds and Critical Value when $\none=\ntwo$}
\label{sec_bounds}
In this section we establish upper and lower bounds  on the critical value $p^*$ of the phase transition. Following the derivation in Appendix \ref{appen_bound}, in Case 2 we have,  almost surely,
\begin{align}
\label{eqn_critical_value_UB_main}
\frac{\lambda_2(\mathbf{L})}{n}
& \leq \frac{p}{2}+ \frac{|\none-\ntwo|p}{2n}  \nonumber \\
&~~~+ \frac{\lambda_2(\bL_1) +\lambda_2(\bL_2)-\left| \lambda_2(\bL_1) -\lambda_2(\bL_2) \right|}{2n},
\end{align}
and
\begin{align}
\label{eqn_critical_value_LB_main}
\frac{\lambda_2(\bL)}{n} &\geq 
 \frac{p}{2}-\frac{|\none-\ntwo| p}{2n} \nonumber \\
&~~~+ \frac{\lambda_2(\bL_1) +\lambda_2(\bL_2)-\left| \lambda_2(\bL_1) -\lambda_2(\bL_2) \right|}{2n}.
\end{align}
Let $p^*$ be the critical value of the phase transition in Case 1 to Case 2. There is a phase transition on the asymptotic value of $\frac{\lambda_2(\bL)}{n}$ since the slope of $\frac{\lambda_2(\bL)}{n}$ converges to 1 almost surely when $p \leq p^*$, whereas from (\ref{eqn_critical_value_UB_main}) $\frac{\lambda_2(\bL)}{n}-p \leq \frac{\lb |\none-\ntwo|-n \rb p}{2n} + \frac{\lambda_2(\bL_1) +\lambda_2(\bL_2)-\left| \lambda_2(\bL_1) -\lambda_2(\bL_2) \right|}{2n}$ when $p \geq p^*$.
Substituting $p^*$ into (\ref{eqn_critical_value_UB_main}), we obtain an asymptotic upper bound $\pUB$ on the critical value $p^*$, where 
\begin{align}
\label{p_UB_main}
\pUB  = \frac{\lambda_2(\bL_1) +\lambda_2(\bL_2)-\left| \lambda_2(\bL_1) -\lambda_2(\bL_2) \right|}{n-|\none-\ntwo|}.
\end{align}
Similarly, by substituting $p^*$ into (\ref{eqn_critical_value_LB_main}),  we obtain an asymptotic lower bound $\pLB$, where 
\begin{align}
\label{p_LB_main}
\pLB = \frac{\lambda_2(\bL_1) +\lambda_2(\bL_2)-\left| \lambda_2(\bL_1) -\lambda_2(\bL_2) \right|}{n+|\none-\ntwo|}.
\end{align}
Comparing (\ref{p_UB_main}) with (\ref{p_LB_main}), the gap between $\pUB$ and $\pLB$ is
\begin{align}
\label{label_eqn_gap}
\frac{|\none-\ntwo|}{2\none \ntwo} \lb \lambda_2(\bL_1) +\lambda_2(\bL_2)-\left| \lambda_2(\bL_1) -\lambda_2(\bL_2) \right| \rb.
\end{align}
Note that when $\none=\ntwo$, the equality in (\ref{eqn_critical_value_LB_2}) holds and the gap in (\ref{label_eqn_gap}) vanishes. This means in Case 2 when $\none=\ntwo$,
\begin{align}
\frac{\lambda_2(\bL)}{n} &\asconv \frac{p}{2} +\frac{\lambda_2(\bL_1) +\lambda_2(\bL_2)-\left| \lambda_2(\bL_1) -\lambda_2(\bL_2) \right|}{2n} \nonumber \\
&=:\frac{p}{2} + c^*,
\end{align}
where $c^*=\frac{\lambda_2(\bL_1) +\lambda_2(\bL_2)-\left| \lambda_2(\bL_1) -\lambda_2(\bL_2) \right|}{2n}$, and
the critical value
\begin{align}
\label{eqn_critical_value_equal_size}
p^* \asconv \frac{\lambda_2(\bL_1) +\lambda_2(\bL_2)-\left| \lambda_2(\bL_1) -\lambda_2(\bL_2) \right|}{n}.
\end{align}

The bounds and the critical value $p^*$ can be specified for some special types of graphs.\\
\begin{itemize}
  \item \textbf{Complete graph:} when each subnetwork is a complete graph (i.e., a clique), $\lambda_j(\bL_i)=n_i$ for all $j \geq 2$ \cite{Mieghem10}. Therefore $\pUB=1$ and $\pLB=\frac{1+c-|1-c|}{1+c+|1-c|}$, where $\frac{\none}{\ntwo} \ra c >0$.
When $n_1=n_2$, $p^* \asconv 1$. This result coincides with the intuition 
that communities that are completely connected are the most detectable.
  \item \textbf{Star graph:} when each subnetwork is a star graph, $\lambda_2(\bL_i)=1$ \cite{Mieghem10}. Since $\pUB = 0$ for all $\none$, $\ntwo$ such that $\frac{\none}{\ntwo} \ra c >0$, we have $p^* \asconv 0$. This means that spectral clustering can not correctly identify the network if each subnetwork is a star graph.
  \item \textbf{Stochastic block model:} when each subnetwork is generated by the Erdos-Renyi graph with edge connection probability $p_i$,
$\lambda_2 \lb \frac{\bL_i}{n_i} \rb \asconv p_i$. This is proved in Appendix \ref{appen_ER}. Therefore $\pUB=\frac{c \pone + \ptwo-|c \pone-\ptwo|}{1+c-|1-c|}$ and $\pLB=\frac{c \pone + \ptwo-|c \pone - \ptwo|}{1+c+|1-c|}$. When $\none = \ntwo$, the critical value
   $p^* \asconv \frac{\pone + \ptwo -|\pone - \ptwo|}{2}$.
\end{itemize}

For community detection with multiple (more than two) subnetworks, we can use successive spectral clustering on the discovered subnetworks.
Assume there are $M$ arbitrarily connected subnetworks with Bernoulli-type random interconnections between subnetworks.
Let $\mathcal{I}$ denote a subset of indices $\{1,2,\ldots,M\}$ such that $\mathcal{I}$ and its set complement $-\mathcal{I}$ are nonempty, and the two corresponding aggregated subnetworks composed of subnetworks indexed by $\mathcal{I}$ and $-\mathcal{I}$ are connected respectively.
Let $\bL_\mathcal{I}$ denote the graph Laplacian matrix of the connected aggregated subnetwork from $\mathcal{I}$ and let $\bL_{-\mathcal{I}}$ denote the graph Laplacian matrix of the connected aggregated subnetwork from $-\mathcal{I}$. Let $n_\mathcal{I}$ and $n_{-\mathcal{I}}$ denote the corresponding aggregated subnetwork size. Then, following the previous derivations, the asymptotic phase transition bounds are
\begin{align}
\label{bound_multi_community}
\pUB  = \min_{\mathcal{I}\subset\{1,2,\ldots,M\}}\frac{\lambda_2(\bL_\mathcal{I}) +\lambda_2(\bL_{-\mathcal{I}})-\left| \lambda_2(\bL_\mathcal{I}) -\lambda_2(\bL_{-\mathcal{I}}) \right|}{n-|n_\mathcal{I}-n_{-\mathcal{I}}|};  \\
\pLB = \min_{\mathcal{I}\subset\{1,2,\ldots,M\}}\frac{\lambda_2(\bL_\mathcal{I}) +\lambda_2(\bL_{-\mathcal{I}})-\left| \lambda_2(\bL_\mathcal{I}) -\lambda_2(\bL_{-\mathcal{I}}) \right|}{n+|n_\mathcal{I}-n_{-\mathcal{I}}|}.
\end{align}
That is, the phase transition bounds are determined by the connected aggregated subnetwork that is the least separable from other subnetworks.

\section{Numerical Experiments}
\label{Sec_performance}
\subsection{Validation of phase transition theory on simulated networks}

\begin{figure}[t]
	\begin{minipage}{.5\textwidth}
		\includegraphics[width=3.8in]{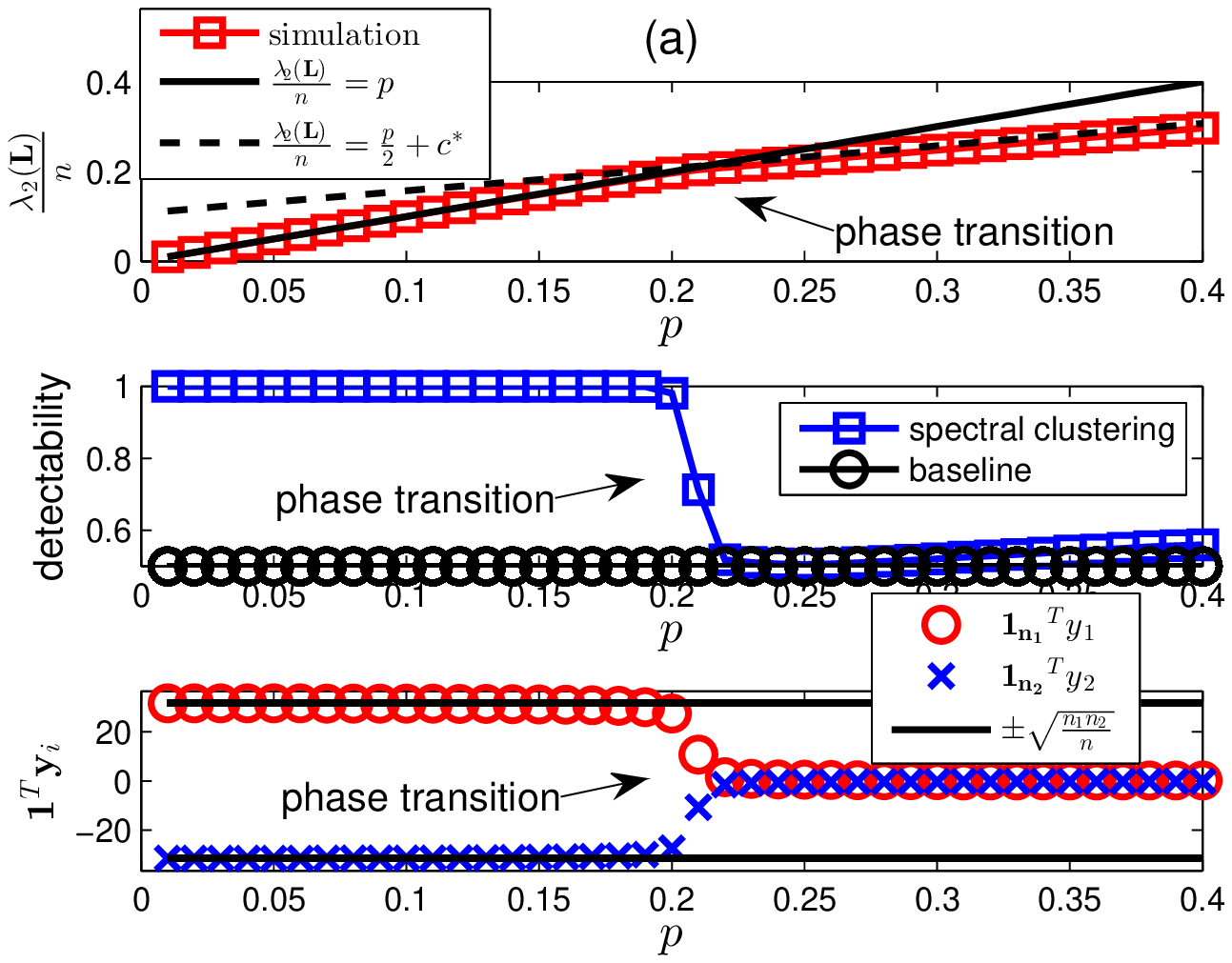}
	\end{minipage}
	\begin{minipage}{.5\textwidth}
		\includegraphics[width=3.8in]{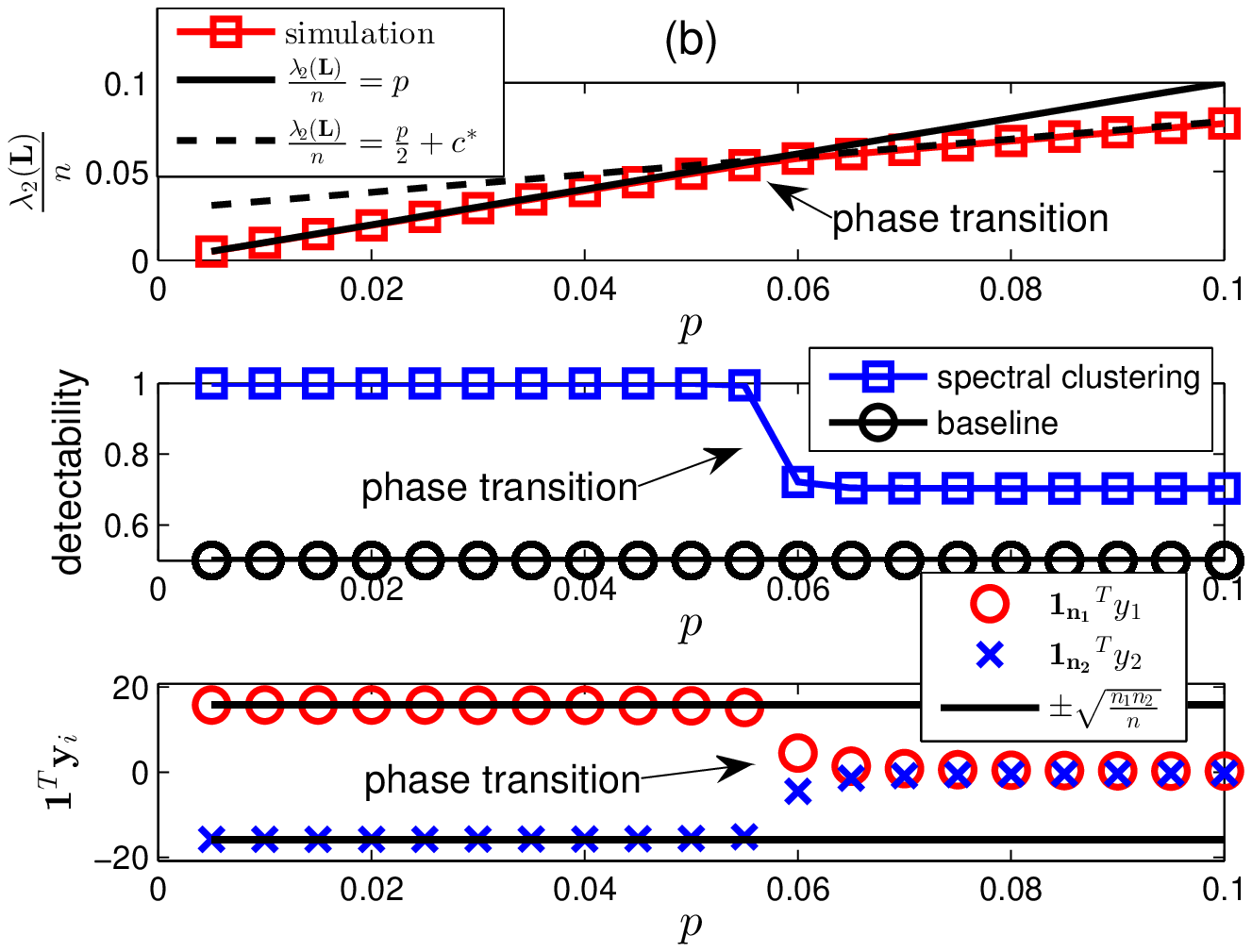}
	\end{minipage}
	\caption{(a) Two identical Erdos-Renyi subnetworks. $n_1$=2000, $n_2$=2000, $p_1$=0.25 and $p_2$=0.25. The empirical critical value $p^*$=0.2142. (b) Two small-world subnetworks. $n_1$=500 and $n_2$=500. Each subnetwork is generated by the Watts-Strogatz small-world network model \cite{Watts98} with average degree 100. The edge rewiring probabilities are 0.2 and 0.8, respectively.
		The empirical critical value $p^*$=0.0566.  The simulation results are averaged over 500 runs and they validate the phase transition analysis.}
	\label{Fig_simulated_network}
\end{figure}

For community detection on simulated networks, the network detectability is defined as the fraction of nodes that are correctly identified.
If the network sizes $n_1$ and $n_2$ are known a priori, a naive identification strategy is to assign all nodes to the subnetwork that has larger network size.
The detectability of the naive strategy, $\max \{\frac{\none}{n},\frac{\ntwo}{n}\}$, is referred to as the baseline detection performance.
Fig. \ref{Fig_simulated_network} (a) shows the case that the two subnetworks are generated by Erdos-Renyi (ER) graphs with the same network sizes and connection probabilities (i.e., the conventional stochastic block model setting that $\none=\ntwo$ and $\pone=\ptwo$). The empirical critical value is $p^*$=0.2142. Note that following the derivations for the stochastic block model in Sec. \ref{sec_bounds}, the empirical value of $p^*$ will converge to $0.25$ as we increase $n$.
The simulation results verify the phase transition effect that $\frac{\lambda_2(\bL)}{n}$ approaches $p$ when $p \leq p^*$ and  $\frac{\lambda_2(\bL)}{n}$ approaches $\frac{p}{2}+c^*$ when $p \geq p^*$, where $c^*=\frac{\lambda_2(\bL_1) +\lambda_2(\bL_2)-\left| \lambda_2(\bL_1) -\lambda_2(\bL_2) \right|}{2n}$. Moreover, the community detection performance transitions from almost perfect detectability to low detectability at $p^*$. As derived in (\ref{eqn_Lagrange17}), the Fiedler vector components $\yone$ and $\ytwo$ are constant vectors with opposite signs for $p$ below $p^*$, and $\onenone^T \yone \ra 0$ and $\onentwo^T \ytwo \ra 0$ above phase transition. Similar results are shown in Fig. \ref{Fig_simulated_network} (b), where the two subnetworks are generated by the Watts-Strogatz small-world network model \cite{Watts98} with the same average degree and different edge rewiring probabilities. The empirical critical value of this network is $p^*$=0.0566.
The low critical value of the Watts-Strogatz small-world network model can be explained by the fact that given the same number of nodes and edges, the algebraic connectivity of such a small-world network increases as the edge rewiring probability increases \cite{Olfati-Saber07}. When the edge rewiring probability is equal to one, the Watts-Strogatz network is equivalent to a Erdos-Renyi graph.

\begin{figure}[t]
 \begin{minipage}{.5\textwidth}
  \includegraphics[width=3.6in]{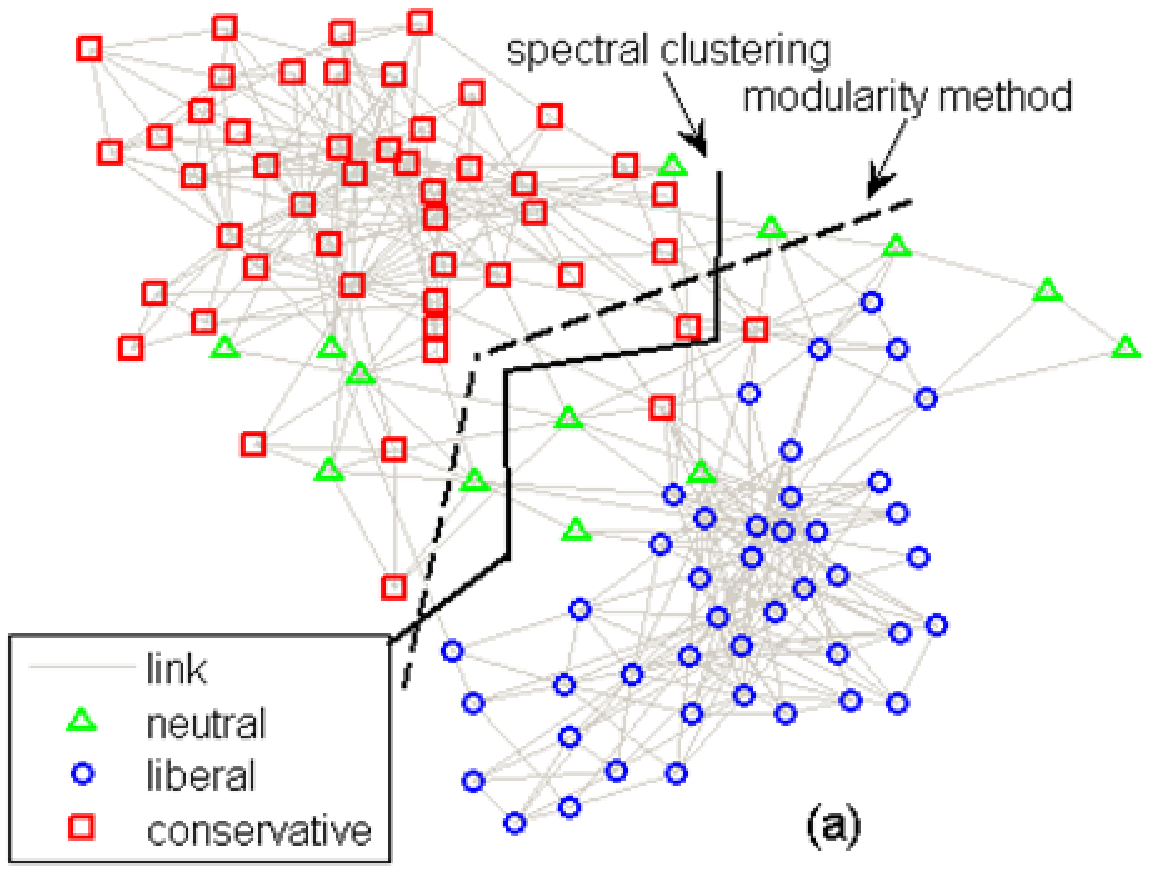}
 \end{minipage}
 \begin{minipage}{.5\textwidth}
  \includegraphics[width=3.6in]{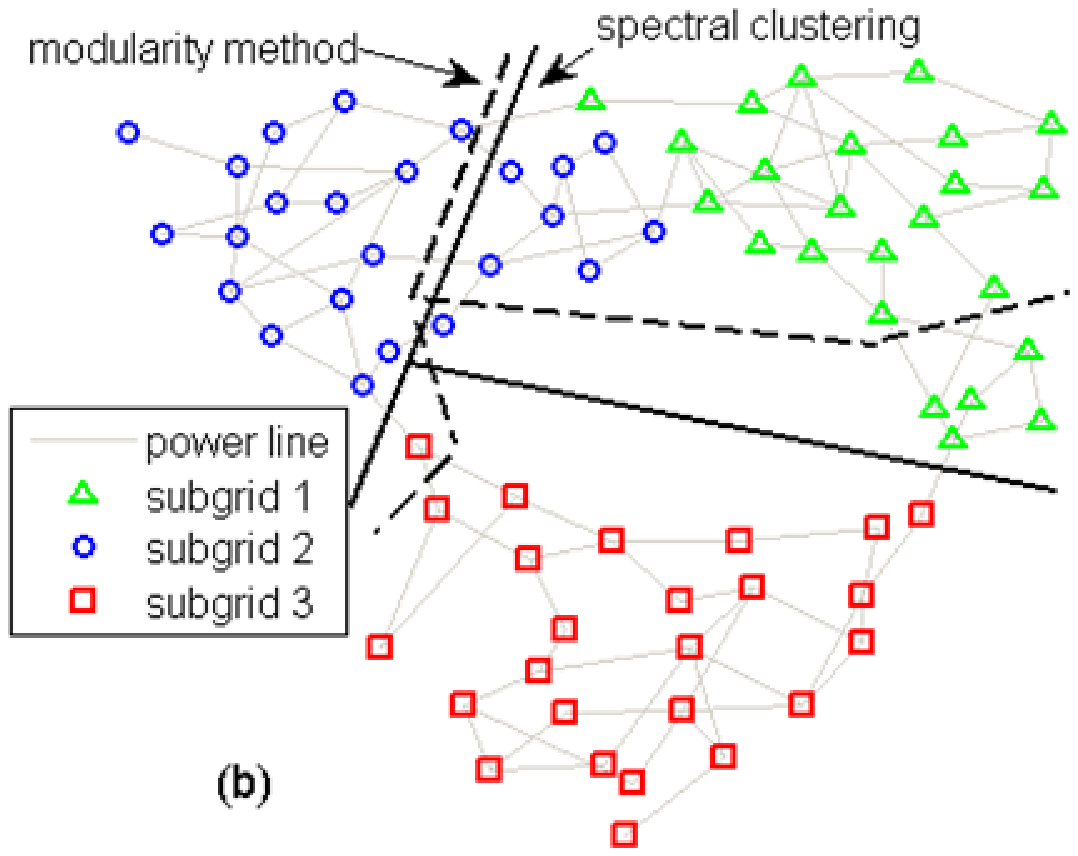}
 \end{minipage}
 \caption{ (a) Co-purchasement of political books on Amazon \cite{Newman06PNAS}. Nodes are political books and edges are co-purchasements. Neglecting the 13 books with neutral labels, 3 books are misidentified by the modularity method and 2 books are misidentified by spectral clustering. (b) IEEE reliability test system consisting of 3 subsystems \cite{Grigg99}. Nodes are power stations and edges are power lines.
 The first cut via spectral clustering perfectly separates subgrid 3 from subgrids 1 and 2.
 Overall, 14 power stations are misidentified by the modularity method and 8 power stations are misidentified by spectral clustering.}
   \label{Fig_network_data}
\end{figure}

\subsection{Application to establishing the phase transition for real-world network data}
Based on the phase transition results in Sec. \ref{sec_bounds}, we propose an empirical method to assess the reliability of discovered communities. In this method 
we explicitly estimate the phase transition bounds $\pUB$, $\pLB$ and the external edge connection probability $p$ from the data.
Let $\widehat{\mathbf{L}}_i$ be the graph Laplacian matrix of the identified subnetwork $i$ having network size $\widehat{n}_i$. Using (\ref{p_UB}) and (\ref{p_LB}), the empirical estimators of these parameters are
\begin{align}
\label{eqn_emp_hpLb}
\hpLB&=\frac{\lambda_2(\hLone) +\lambda_2(\hLtwo)-\left| \lambda_2(\hLone) -\lambda_2(\hLtwo) \right|}{n+|\hnone-\hntwo|}; \\
\label{eqn_emp_hpUb}
\hpUB&=\frac{\lambda_2(\hLone) +\lambda_2(\hLtwo)-\left| \lambda_2(\hLone) -\lambda_2(\hLtwo) \right|}{n-|\hnone-\hntwo|}; \\
\label{eqn_emp_hp}
\widehat{p}&=\text{number of identified external edges}/\hnone \hntwo.
\end{align}
Based on these spectral estimates, the performance of community detection can be classified into three categories. If $\widehat{p} \leq \hpLB$, the network is in the reliable detection region. If $\hpLB < \widehat{p} < \hpUB$, the network is in the intermediate detection region. If $\widehat{p} \geq \hpUB$, the network is in the unreliable detection region.
The network descriptions and the results of spectral clustering and the modularity method \cite{Newman06PNAS} are summarized in Table \ref{Table_network_data}.
Note that no information beyond the network topology is used to estimate these phase transition parameters. The community labels in Table \ref{Table_network_data} are used to verify the network detectability.

We illustrate this method on two datasets. 
The first dataset is the co-purchasement data between 105 American political books sold on Amazon \cite{Newman06PNAS}. An edge exists between two books if they are frequently purchased by the same buyer. Three labels, \emph{liberal}, \emph{conservative} and \emph{neutral}, are specified in \cite{Newman06PNAS}. We perform community detection by separating the books into two groups since there are only 13 books with neutral labels. The graph cuts identified by spectral clustering and the modularity method are shown in Fig. \ref{Fig_network_data} (a). Neglecting the 13 books with neutral labels, 3 books are misidentified by the modularity method and 2 books are  misidentified by spectral clustering. The empirical estimate (\ref{eqn_emp_hp}) of the external edge connection probability is $\widehat{p}$=0.0073 and the empirical estimate (\ref{eqn_emp_hpLb}) of the lower bound on the phase transition is $\hpLB$=0.0127. The fact that $\widehat{p}<\hpLB$ provides evidence that these communities are in fact detectable, providing reassure about their validity.

\begin{table*}[t]
\caption{Data descriptions, spectral estimates of phase transition parameters, and network detectability. These results suggest that we can use the phase transition estimates to experimentally validate estimates obtained from the community detection procedure. $\widehat{p} \leq \hpLB$, $\hpLB < \widehat{p} < \hpUB$, and $\widehat{p} \geq \hpUB$ correspond to the reliable, intermediate, and unreliable detection regions, respectively.}
\label{Table_network_data}
\begin{center}
\begin{tabular}{lllllll}
& & & \multicolumn{1}{ c| }{\textbf{Spectral Estimates}}  & \multicolumn{3}{ c }{\textbf{Network Detectability}} \\  \cline{4-7}
\multicolumn{1}{c}{\bf Network}  &\multicolumn{1}{c}{\bf Class} &\multicolumn{1}{c}{\bf Nodes~/~\bf Edges}  & \multicolumn{1}{c}{$\hpLB~/~\hpUB~/~\widehat{p}$}
&\multicolumn{1}{c}{\bf {Mod. }} &\multicolumn{1}{c}{\bf Spec.} &\multicolumn{1}{c}{\bf Oracle}
\\ \hline  \\
Political books             &2 &105 / 441 &.0127 / .013 ~~/ .0073& .8476 & .8571 & .8762\\
IEEE RTS (1st cut)      &3  &73 ~~/ 108 &.0016 / .003 ~~/ .0017 & .9041 & 1 & 1\\
IEEE RTS (2nd cut)       &3  &73 ~~/ 108 &.003~~~/ .0047 / .0078 & .8082 & .8904 & 1\\ \hline
\end{tabular}
\end{center}
\end{table*}

The second dataset considered is the IEEE reliability test systems (RTS) for power system \cite{Grigg99}. The network consists of 3 interconnected subsystems. Community detection is performed by first partitioning the network into two subnetworks and then repartitioning the largest subnetwork. The graph cuts are shown in Fig. \ref{Fig_network_data} (b). Note that the first cut via spectral clustering perfectly separates subgrid 3 from subgrids 1 and 2. This is consistent with the fact that the empirically estimated value (\ref{eqn_emp_hp}) is  $\widehat{p}$=0.0017, which is close to the estimated phase transition lower bound $\hpLB$=0.0016 and $\widehat{p}<\hpUB$. For the second cut on the subnetwork consisting of subgrids 1 and 2, 8 power stations are misidentified by spectral clustering, which is consistent with the empirical finding that $\widehat{p}>\hpUB$. The fact that this second cut discovered communities are above the phase transition threshold might explain why 14 power stations are misidentified by the modularity method.
These results suggest that we can use the proposed phase transition estimates to experimentally validate
estimates obtained from the community detection procedure.

\section{Conclusion}
In this paper, we establish and quantify a phase transition threshold for spectral clustering based community detection.
The critical value of this phase transition is a function of the probability of an edge connecting two subnetworks. Bounds on the critical value $p^*$ are derived and validated by simulation. The bounds are tight when the two subnetwork sizes are identical.
We use real-world network data to show that these phase transition bounds can be estimated to validate the detection reliability of spectral community detection methods.

\section{Appendix}
\subsection{Proof of (\ref{eqn_Latala})}
\label{appen_Latala}
Since $\bDelta=\bC-\bCbar$, we have $\bDelta_{ij}=1-p$ with probability $p$ and $\bDelta_{ij}=-p$ with probability $1-p$.
Latala's theorem \cite{Latala05} states that for any random matrix $\mathbf{M}$ with statistically independent and zero mean entries, there exists a positive constant $c_1$ such that
\begin{align}
\mathbb{E} \Lb \sigma_1(\mathbf{M})\Rb &\leq c_1 \lb \max_i\sqrt{\sum_j \mathbb{E} \Lb \mathbf{M}_{ij}^2 \Rb}+\max_j\sqrt{\sum_i \mathbb{E} \Lb \mathbf{M}_{ij}^2 \Rb} \right. \nonumber \\
&~~~\left.+ \sqrt[4]{\sum_{ij} \mathbb{E} \Lb \mathbf{M}_{ij}^4 \Rb}\rb.
\end{align}
It is clear that $\mathbb{E} \Lb \bDelta_{ij} \Rb=0$ and each entry in $\bDelta$ is independent. By using $\mathbf{M}=\frac{\bDelta}{\sqrt{\none \ntwo}}$ in Latala's theorem, since $p \in [0,1]$, we have $\max_i\sqrt{\sum_j \mathbb{E} \Lb \mathbf{M}_{ij}^2 \Rb}=O(\frac{1}{\sqrt{n_1}})$, $\max_j\sqrt{\sum_i \mathbb{E} \Lb \mathbf{M}_{ij}^2 \Rb}=O(\frac{1}{\sqrt{n_2}})$, and $\sqrt[4]{\sum_{ij} \mathbb{E} \Lb \mathbf{M}_{ij}^4 \Rb}=O(\frac{1}{\sqrt[4]{\none \ntwo}})$.
Therefore $\mathbb{E} \Lb \sigma_1\lb \frac{\bDelta}{\sqrt{\none \ntwo}} \rb \Rb \ra 0$ as $\none \ra \infty$ and $\ntwo \ra \infty$.

\subsection{Proof of (\ref{eqn_Talagrand})}
\label{appen_Talagrand}
Talagrand's concentration inequality is stated as follows. Let $f: \mathbb{R}^k \mapsto \mathbb{R}$ be a convex and 1-Lipschitz function.
Let $\bx \in \mathbb{R}^k$ be a random vector and assume that every element of $\bx$ satisfies
$|\bx_i|  \leq K$ for all $i=1,2,\ldots,k$, with probability one.
Then there exist positive constants $c_2$ and $c_3$ such that $\forall \epsilon >0$,
\begin{align}
\text{Pr}\lb \left| f(\bx)-\mathbb{E} \Lb f(\bx)  \Rb \right| \geq \epsilon\rb \leq c_2 \exp \lb \frac{-c_3 \epsilon^2}{K^2} \rb.
\end{align}
It is well-known that the largest singular value of a matrix $\mathbf{M}$ can be represented as $\sigma_1(\mathbf{M})=\max_{\bz^T\bz=1}||\mathbf{M} \bz||_2$ \cite{HornMatrixAnalysis} such that $\sigma_1(\mathbf{M})$ is a convex and 1-Lipschitz function. Therefore applying Talagrand's inequality by substituting $\mathbf{M}=\frac{\bDelta}{\sqrt{\none \ntwo}}$ and using the facts that $\mathbb{E} \Lb \sigma_1\lb \frac{\bDelta}{\sqrt{\none \ntwo}} \rb \Rb \ra 0$ and $\frac{\bDelta_{ij}}{\sqrt{\none \ntwo}} \leq \frac{1}{\sqrt{\none \ntwo}}$, we have
\begin{align}
\text{Pr}\lb  \sigma_1 \lb \frac{\bDelta}{\sqrt{\none \ntwo}} \rb \geq \epsilon \rb \leq c_2 \exp \lb -c_3 \none \ntwo \epsilon^2 \rb.
\end{align}
Note that, since for any positive integer $\none, \ntwo >0$  $\none \ntwo \geq \frac{\none+\ntwo}{2}$,
$\sum_{\none,\ntwo} c_2 \exp \lb -c_3 \none \ntwo \epsilon^2 \rb < \infty$.
Hence, by Borel-Cantelli lemma \cite{Resnick13},
$\sigma_1 \lb \frac{\bDelta}{\sqrt{\none \ntwo}} \rb \asconv 0$
when $\none, \ntwo \ra \infty$.
Finally, a standard matrix perturbation theory result \cite{HornMatrixAnalysis} is
$|\sigma_i(\bCbar+\bDelta)-\sigma_i(\bCbar)| \leq \sigma_1(\bDelta)$
for all $i$, and as $\sigma_1\lb \frac{\bDelta}{\sqrt{\none \ntwo}} \rb \asconv 0$, we have
\begin{align}
&\sigma_1\lb \frac{\bC}{\sqrt{\none \ntwo}} \rb=\sigma_1\lb \frac{\bCbar+\bDelta}{\sqrt{\none \ntwo}} \rb \asconv \sigma_1\lb \frac{\bCbar}{\sqrt{\none \ntwo}} \rb=p; \\
&\sigma_i \lb \frac{\bC}{\sqrt{\none \ntwo}} \rb \asconv 0~~\forall i \geq 2
\end{align}
when $\none \ra \infty$ and $\ntwo \ra \infty$.

\subsection{Proof of $\lambda_2 \lb \frac{\bL_i}{n_i} \rb \asconv p_i$ for the stochastic block model}
\label{appen_ER}
Consider a network with adjacency matrix $\bA$ and size $n$ generated by the Erdos-Renyi graph with edge connection probability $q$. Each entry of $\bA$ is an i.i.d Bernoulli random variable with connection probability $q$. Write the graph Laplacian matrix as $\bL=\bD-\bA=\bDbar-\bAbar+ \bDelta \bD - \bDelta \bA$, where $\bDelta \bD = \bD-\bDbar$, $\bDelta \bA= \bA-\bAbar$, $\bAbar=q \bone_{n} \bone_{n}^T$ and $\bDbar=\text{diag}(n q,\ldots,n q)$. Following the arguments in Appendix \ref{appen_Latala} and \ref{appen_Talagrand}, since $\bDelta \bA_{ij}=1-q$ with probability $q$ and $\bDelta \bA_{ij}=-q$ with probability $1-q$,
$\sigma_1 \lb \frac{\bDelta \bA}{n} \rb \asconv 0$ when $n \ra \infty$. Let $\mathfrak{B}_{s,q}$ be a binomial random variable which is the sum of $s$ i.i.d Bernoulli random variables with success probability $q$. We have $\bDelta \bD_{ij}=\mathfrak{B}_{n,q}-n q$ if $i=j$ and $\bDelta \bD_{ij}=0$ otherwise.
By Bernstein's concentration inequality \cite{Resnick13}, for any $\epsilon >0$, there exists a positive constant $c_4$ such that
\begin{align}
\text{Pr} \lb \left| \frac{\bDelta \bD_{ii}}{n} \right| \geq \epsilon \rb \leq \exp{\lb -c_4 n \epsilon^2 \rb}~\forall i.
\end{align}
Since $\bDelta \bD$ is a diagonal matrix, $\sigma_1 \lb \frac{\bDelta \bD}{n} \rb=\max_{i} \left| \frac{\bDelta \bD_{ii}}{n} \right|$. Using the union bound, for any $\epsilon>0$,
\begin{align}
\label{eqn_ER_Bernstein}
\text{Pr} \lb \sigma_1 \lb \frac{\bDelta \bD}{n} \rb \geq \epsilon  \rb &\leq \sum_{i=1}^n \text{Pr} \lb \left| \frac{\bDelta \bD_{ii}}{n} \right| \geq \epsilon \rb \nonumber \\
&\leq n\exp{\lb -c_4 n \epsilon^2 \rb}.
\end{align}
Since $\sum_n n\exp{\lb -c_4 n \epsilon^2 \rb} < \infty$, applying Borel-Cantelli lemma gives $\sigma_1 \lb \frac{\bDelta \bD}{n} \rb \asconv 0$.
Using the standard matrix perturbation theory result \cite{HornMatrixAnalysis},
$\left|\sigma_i \lb \frac{\bDbar-\bAbar+ \bDelta \bD - \bDelta \bA}{n} \rb-\sigma_i \lb \frac{\bDbar-\bAbar}{n} \rb \right| \leq \sigma_1 \lb \frac{\bDelta \bD-\bDelta \bA}{n} \rb$ for all $i$. By the fact that $\sigma_1 \lb \frac{\bDelta \bD-\bDelta \bA}{n} \rb \leq \sigma_1 \lb \frac{\bDelta \bD}{n} \rb+ \sigma_1 \lb \frac{\bDelta \bA}{n} \rb \asconv 0$, we have
\begin{align}
\sigma_i \lb \frac{\bL}{n} \rb \asconv   \sigma_i \lb \frac{\bDbar-\bAbar}{n} \rb
\end{align}
for all $i$, and $\sigma_i \lb \frac{\bL}{n} \rb=\lambda_i \lb \frac{\bL}{n} \rb$ since $\bL$ is a PSD square matrix. Finally, since $\bDbar-\bAbar$ is the graph Laplacian matrix of a complete graph with edge weight $q$, we have $\lambda_2 \lb \frac{\bL}{n} \rb \asconv q$.

\subsection{Proof of upper and lower bounds on $p^*$}
\label{appen_bound}
From (\ref{eqn_alge}) and (\ref{eqn_Laplacian_block}) we know that
\begin{align}
\label{eqn_alge_express}
\lambda_2(\mathbf{L})=\yone^T (\Lone+\Done) \yone + \ytwo^T (\Ltwo+\Dtwo) \ytwo -2\yone^T \bC \ytwo
\end{align}
subject to $\onenone^T\yone+\onentwo^T\ytwo=0$ and $\yone^T\yone+\ytwo^T\ytwo=1$.
In Case 2, since $\onenone^T \yone \ra 0$ and $\onentwo^T \ytwo \ra 0$ almost surely, recalling the definition $\bDelta=\bC-\bCbar$,
\begin{align}
\frac{1}{\sqrt{\none \ntwo}} \yone^T\bC \ytwo&= \frac{1}{\sqrt{\none \ntwo}}\yone^T \lb \bCbar+\bDelta \rb\ytwo \nonumber \\
&=\frac{1}{\sqrt{\none \ntwo}} \lb \yone^T  \bCbar\ytwo + \yone^T \bDelta \ytwo \rb \nonumber \\
& \leq \frac{1}{\sqrt{\none \ntwo}} \lb \yone^T  \bCbar\ytwo + \| \yone \|_2 \| \ytwo \|_2 \cdot \sigma_1(\bDelta) \rb \nonumber \\
&\asconv 0
\end{align}
by the fact that $\sigma_1\lb \frac{\bDelta}{\sqrt{\none \ntwo}} \rb \asconv 0$ in Appendix \ref{appen_Talagrand} and $\bCbar=p \onenone \onentwo^T$. Furthermore, by the facts that $\Done=\text{diag}\left(\bC\onentwo\right)$ and $\Dtwo=\text{diag}\left(\bC^T\onenone\right)$, (\ref{eqn_Talagrand}) gives
\begin{align}
\frac{1}{\ntwo}\yone^T \Done \yone \asconv  p \yone^T \yone;~~
\frac{1}{\none}\yone^T \Dtwo \yone \asconv  p \ytwo^T \ytwo.
\end{align}
Therefore in Case 2 we have
\begin{align}
\label{eqn_case2_alge}
&\frac{\lambda_2(\mathbf{L})}{n} \asconv \nonumber \\
&\min_{\bx \in \mathcal{S}} \LB \frac{1}{n} \lb \xone^T \Lone \xone + \xtwo^T \Ltwo \xtwo+ \ntwo p \xone^T \xone
+ \none p \xtwo^T \xtwo \rb \RB,
\end{align}
where
\begin{align}
&\mathcal{S}=\LB \bx=[\xone~\xtwo]^T, \xone \in \mathbb{R}^\none, \xtwo \in \mathbb{R}^\ntwo: \right. \nonumber \\
&~~~~~~~\left.\onenone^T\xone=\onentwo^T\xtwo=0,~\xone^T\xone+\xtwo^T\xtwo=1 \RB.
\end{align}
Define two sets
\begin{align}
&\mathcal{S}_1=\LB \bx=[\xone~\xtwo]^T, \xone \in \mathbb{R}^\none, \xtwo \in \mathbb{R}^\ntwo: \right. \nonumber \\
&~~~~~~~~\left.\onenone^T\xone=\onentwo^T\xtwo=0,~\xone^T\xone=1,~\xtwo^T\xtwo=0 \RB;  \\
&\mathcal{S}_2=\LB \bx=[\xone~\xtwo]^T, \xone \in \mathbb{R}^\none, \xtwo \in \mathbb{R}^\ntwo: \right. \nonumber \\
&~~~~~~~~\left. \onenone^T\xone=\onentwo^T\xtwo=0,~\xone^T\xone=0,~\xtwo^T\xtwo=1 \RB,
\end{align}
and define
\begin{align}
&\mu_i(\bL)=  \nonumber \\
&\min_{\bx \in \mathcal{S}_i} \LB \frac{1}{n} \lb \xone^T \Lone \xone + \xtwo^T \Ltwo \xtwo + \ntwo p \xone^T \xone + \none p \xtwo^T \xtwo \rb \RB.
\end{align}

Since $\mathcal{S}_1,\mathcal{S}_2 \subseteq \mathcal{S}$, we have, almost surely,
\begin{align}
\frac{\lambda_2(\mathbf{L})}{n}& \leq \min \LB  \mu_1(\bL) , \mu_2(\bL)\RB \nonumber \\
& = \min \LB  \frac{\lambda_2(\bL_1) + \ntwo p}{n} , \frac{\lambda_2(\bL_2)+\none p}{n}\RB \nonumber \\
& = \frac{p}{2}+\frac{\lambda_2(\bL_1) +\lambda_2(\bL_2)}{2n} \nonumber \\
&~~~-\frac{\left| \lambda_2(\bL_1) -\lambda_2(\bL_2)+(\ntwo-\none) p \right|}{2n} \nonumber \\
& \leq \frac{p}{2}+ \frac{|\none-\ntwo|p}{2n}  \nonumber \\
&~~~+ \frac{\lambda_2(\bL_1) +\lambda_2(\bL_2)-\left| \lambda_2(\bL_1) -\lambda_2(\bL_2) \right|}{2n},
\label{eqn_critical_value_UB}
\end{align}
where we use the facts that $\min\{a,b\}=\frac{a+b-|a-b|}{2}$ and $|a-b| \geq |a|-|b|$. Note that the equality in (\ref{eqn_critical_value_UB}) holds if $\none=\ntwo$.  Let $p^*$ be the critical value for phase transition from Case 1 to Case 2. There is a phase transition on the asymptotic value of $\frac{\lambda_2(\bL)}{n}$ since the slope of $\frac{\lambda_2(\bL)}{n}$ converges to 1 almost surely when $p \leq p^*$, whereas from (\ref{eqn_critical_value_UB}) $\frac{\lambda_2(\bL)}{n}-p \leq \frac{\lb |\none-\ntwo|-n \rb p}{2n} + \frac{\lambda_2(\bL_1) +\lambda_2(\bL_2)-\left| \lambda_2(\bL_1) -\lambda_2(\bL_2) \right|}{2n}$ when $p \geq p^*$.
From (\ref{eqn_Lagrange11}), we obtain an asymptotic upper bound $\pUB$ on the critical value $p^*$ by substituting $p^*$ into (\ref{eqn_critical_value_UB}).
\begin{align}
\label{p_UB}
\pUB  = \frac{\lambda_2(\bL_1) +\lambda_2(\bL_2)-\left| \lambda_2(\bL_1) -\lambda_2(\bL_2) \right|}{n-|\none-\ntwo|}.
\end{align}

For the lower bound, with (\ref{eqn_case2_alge}) we have that in Case 2,
\begin{align}
\frac{\lambda_2(\bL)}{n}& \asconv \min_{\bx \in \mathcal{S}} \LB \frac{1}{n} \lb \xone^T \Lone \xone + \xtwo^T \Ltwo \xtwo \rb \right. \nonumber \\
&\left.~~+ \frac{1}{n} \lb \ntwo p \xone^T \xone + \none p \xtwo^T \xtwo \rb \RB
\nonumber \\
\label{eqn_critical_value_LB_2}
& \geq \min_{\bx \in {S}} \LB \frac{1}{n} \lb \xone^T \Lone \xone + \xtwo^T \Ltwo \xtwo \rb \RB \\
&~~~+ \min_{\bx \in S} \LB \frac{1}{n} \lb \ntwo p \xone^T \xone + \none p \xtwo^T \xtwo \rb \RB \nonumber \\
&= \min\LB \frac{\lambda_2(\bL_1)}{n},\frac{\lambda_2(\bL_2)}{n} \RB + \min \LB \frac{\none p}{n}, \frac{\ntwo p}{n}\RB. \nonumber \\
\label{eqn_critical_value_LB}
&= \frac{p}{2}-\frac{|\none-\ntwo| p}{2n} \nonumber \\
&~~~+ \frac{\lambda_2(\bL_1) +\lambda_2(\bL_2)-\left| \lambda_2(\bL_1) -\lambda_2(\bL_2) \right|}{2n}.
\end{align}
Substituting $p^*$ to (\ref{eqn_critical_value_LB}), we obtain an asymptotic lower bound $\pLB$ on the critical value $p^*$.
\begin{align}
\label{p_LB}
\pLB = \frac{\lambda_2(\bL_1) +\lambda_2(\bL_2)-\left| \lambda_2(\bL_1) -\lambda_2(\bL_2) \right|}{n+|\none-\ntwo|}.
\end{align}

\bibliographystyle{IEEEtran}
\bibliography{IEEEabrv,network_identify}
\nocite{*}

\end{document}